\documentclass[11pt]{article}
\usepackage[utf8]{inputenc}
\usepackage[margin=0.9in]{geometry}
\usepackage{graphicx,verbatim}
\usepackage{amsfonts}
\usepackage{amsmath}
\usepackage{amssymb}
\usepackage{hyperref}

\usepackage{caption}
\usepackage{subcaption}
\usepackage{algpseudocode}
\usepackage{algorithm}
\usepackage[noabbrev]{cleveref}
\usepackage{url}

\newcommand{\eps}{\epsilon}
\newcommand{\polylog}{\text{ polylog }}
\newtheorem{example}{Example}
\title{Matrix Multiplication in Quadratic Time and Energy?\\  Towards a Fine-Grained Energy-Centric Church-Turing Thesis}

 \author{Gregory Valiant \\Stanford University}
\begin{document}

\maketitle
\begin{abstract}
    We describe two algorithms for multiplying $n \times n$ matrices using time and energy $\tilde{O}(n^2)$ under basic models of classical physics.  The first algorithm is for multiplying integer-valued matrices, and the second, quite different algorithm, is for Boolean matrix multiplication.  We hope this work inspires a deeper consideration of physically plausible/realizable models of computing that might allow for algorithms which improve upon the runtimes and energy usages suggested by the parallel RAM model in which each operation requires one unit of time and one unit of energy.
\end{abstract}
\section{Introduction}

Suppose you were presented with a black-box that could multiply any $n\times n$ matrices in quadratic time. Would you be surprised?  Not necessarily---the box might simply be able to leverage an amount of parallelism that scales with $n$.  Specifically, you could trivially parallelize the multiplication across $n$ machines, and run each machine for $O(n^2)$ time, resulting in $O(n^3)$ energy usage but only $O(n^2)$ time.  But what if both the runtime and energy usage of the black-box scaled quadratically?  Such a black-box would be surprising if it operated within a computational model where each arithmetic operation requires one unit of energy.  But are there physically realizable models that do not have this property? And if so, what is the algorithmic landscape for such models, and what physical gadgets or properties do they leverage?  Should we expect to be able to obtain significant polynomial improvements simultaneously for runtime and energy usage for fundamental algorithmic primitives like matrix multiplication?  

There are several motivations for considering these questions.  First, energy is one of the most important computational resources, along with time, and space.  Despite this, there is embarrassingly little theoretical work on low-energy computing, and few theoretical models of computation that explicitly consider energy.  Of course, on the practical side there is a frenzy of effort to design highly parallel and energy-efficient hardware and algorithms---and a  proliferation of analog computing components due to their low energy-usage. 
 Still, a more principled effort to  understand how different physical systems and assumptions could be algorithmically leveraged for low-energy computation might serve to guide the development of alternative hardware and architectures.  

From a more conceptual angle, these questions ask whether the conventional wisdom regarding the time and energy complexity of problems is inherent, or simply due to our RAM-centric view of computing, modeled on computers in the von Neumann architecture.  In light of the extended Church-Turing thesis, we do not expect natural or physics-driven computational processes to obtain super-polynomial improvements in terms of time and energy---quantum computing aside.  In terms of the structure of problems within P, however, we \emph{do} know that different computational models give rise to different polynomial runtimes. 
Despite this, there seems to be little investigation of realistic and physically plausible models of computation that result in significant (polynomial) savings in resources over the standard RAM or parallel-RAM models:
\begin{quote}
\emph{    What is a ``fine-grained'' analog of the extended Church-Turing thesis that takes into account both runtime and energy?  Do plausible non-quantum models of computing admit polynomial savings in terms of time and energy over the RAM or parallel-RAM models where each operation takes a unit of energy? If so, how large can these polynomial factors be, and what are the fundamental lower-bounds for natural problems?}
\end{quote}

\section{Related work}
The earliest analog computers were mechanical in nature and were later replaced with electronic analog computers. A good example of an early analog computer was the differential analyser~\cite{bush1931differential} which was used to solve differential equations. Later there were theoretical models developed for studying the power of analog computation that uses a set of elementary operations such as constants, adders, multipliers and integrators~\cite{shannon1941mathematical,pour1974abstract,gracca2003analog}.  The focus in these works is on computability, as opposed to runtime or energy usage.  

Early theoretical work in the study of energy efficient computation was done in the context of reversible computing, initiated by Landauer and Bennet~\cite{landauer1961irreversibility,bennett1973logical,bennett1988notes}. Landauer's principle~\cite{landauer1961irreversibility} states that erasing a single bit of information requires $k_B T \log 2$ energy, where $k_B$ is Boltzmann’s constant and $T$ is the temperature of the surroundings. The motivation for reversible computing is the stipulation that, from a thermodynamic perspective, such erasures are the \emph{only} aspect of computation that inherently requires energy, and hence if a computation is reversible, there is no theoretical lower bound to the energy required.  More recent work in this vein by Demaine et al.~\cite{demaine2016energy} studies this in a more algorithmic context and revists many common algorithmic primitives (including sorting, graph algorithms and data structures) with the goal of implementing them entirely, or mostly, with reversible operations.  

There has also been a line of theoretical work on a different notion of energy complexity (e.g.~\cite{uchizawa2009energy}).  In that work, the energy complexity of a circuit is defined as the maximum over all inputs, of the number of gates that output $1$ (as opposed to a 0).  This definition corresponds to the energy expended in a natural implementation of such a circuit. The key questions are how the energy complexity can be related to traditional parameters of circuits, such as width or depth.

Our algorithms leverage only classical physics.  Of course, quantum algorithms such as Shor's algorithm~\cite{shor1994algorithms} may yield super-polynomial improvements over classical algorithms, both in terms of runtime and energy.  There is also a significant line of work investigating the extent to which restricted models of quantum computation---such as ``linear optics''~\cite{aaronson2011computational}---can yield super-polynomial speedups.  There are also several interesting quantum algorithms, such as Grover's search~\cite{grover96} and recent work on quantum ``spatial search''~\cite{PhysRevLett.129.160502}, which yield only quadratic speedups over their classical analogs.  Given this interest in polynomial speedups, it is certainly worth understanding whether certain types of non-quantum physical systems can give similar sorts of surprising speedups.  We also note that the challenges to realizing quantum computing in a practical sense appear orthogonal to the challenges of realizing the sort of ``physical'' algorithms we present here.

 On the practical side, energy is one of the most important metrics of computational efficiency.  On mobile devices (phones, watches etc), battery life is a paramount concern.  For training deep neural networks and large-scale scientific computing, energy costs are often significant in comparison to the hardware costs and the salaries of the people involved.  This has sparked a large industry of custom hardware, and renewed interest in analog computing.  Particularly in settings that allow for low-precision, analog circuits seem to   offer significant energy savings for certain problems (see, e.g. the very brief survey~\cite{K_ppel_2021}). 
For specific computational primitives, in particular, matrix vector multiplication, there have been a series of empirical papers exploring analog implementations via memristor crossbar circuits~\cite{hu2014memristor,hu2016dot}.  Additionally, there is a promising wave of work on optical/``photonic'' circuits (e.g.~\cite{shen2017deep}), which seem to offer both increased speed and lower energy for tasks such as forward passes on a deep neural network. The emphasis in these works is on the empirical behaviour, not asymptotic or theoretical properties.

\section{Potential Advantages of Physical Algorithms}
Our algorithms will leverage concrete physical systems that evolve under the laws of classical physics.  Before describing these algorithms, we outline three properties of physical systems that could plausibly be employed to yield time and/or energy improvements over the RAM model:
\begin{itemize}
    \item \textbf{Free Parallelism:} The physical world allows for some level of parallelism ``for free'', as multiple physical systems can evolve in parallel. The initialization/setup of these systems may need to be done serially, but their evolution according to the laws of physics occurs in parallel. 
    \item \textbf{Can Tradeoff Time and Energy:}  Under Newtonian mechanics, suppose it takes one unit of energy to move a unit-mass object one unit distance, with the object beginning and ending at rest.  In a frictionless setting, to move the same object one unit distance in $t$ units of time, the total energy is $1/t^2$, since the object needs to be accelerated to velocity $1/t$, and kinetic energy scales with the \emph{square} of the velocity. This ability to tradeoff between time and energy is exploited in the Boolean Matrix Multiplication algorithm of Section~\ref{sec:bool_mat_mult}.  It is worth noting that a similar scaling is observed with over/under-clocking CPUs (though there is only a narrow range of flexibility in clock-speed of current CPUs), though this scaling is due both to increasing the voltage and increased fan speed required to dissipate the heat. 
    \item \textbf{Sublinear Time/Energy Aggregation:} Physical systems allow for many means for adding or computing the OR of $n$ numbers using a sublinear amount of time and/or energy.  1) Diffusion: If the $n$ quantities to be aggregated are presented as $n$ heat sources, arranged on a $\sqrt{n} \times \sqrt{n}$ two dimensional grid of thermally conducting material (thermally insulated from the outside world), then with no additional energy and time $O(n \log (1/\epsilon))$, the heat equation will drive the conducting plate to a uniform temperature to within $\pm \epsilon$.  
    If the $n$ quantities to be aggregated were presented as $n$ heat sources, arranged within a $n^{1/3} \times n^{1/3} \times n^{1/3}$ cube, then the time for diffusion is sublinear:  $O(n^{2/3} \log (1/\epsilon))$. 2) Newtonian mechanics: given $n$ bits, let the $i$th bit be represented as the presence or absence of a unit mass block at location $i$ along a length $n$ friction-less track.  The OR of these bits can be computed by sliding a unit-mass block along the track with some initial velocity, and measuring whether that block is the first block to reach location $n+1$.   If the initial block has velocity $v$ (and hence energy $O(v^2$)), then if the OR is 0 that block will reach the end at time $n/v$. Provided $v < \sqrt{n}$, this provides a smooth tradeoff between sublinear time and sublinear energy.

\end{itemize}

\section{Physical Assumptions}
In this section, we briefly discuss the assumptions underlying the correctness and runtimes/energy usages of our algorithms.  As with any such assumptions, they become unrealistic at some problem scale.  This is similar to the sense in which the RAM model becomes unrealistic at the problem scales for which the time to communicate a bit of information across the memory footprint is non-negligible.

\subsection{Precision and Measurement Accuracy} Our algorithms  leverage the  assumption that physical quantities (e.g. mass, length) of value $b$ can be  measured to accuracy $\pm \epsilon$, using a time and energy cost of $\log(\max(1,b))+\log \frac{1}{\eps}$.  Additionally, the time and energy cost of fabricating a component with desired mass or length $b \pm \epsilon$ is $O(b + \log(1/\epsilon))$. These assumptions are reasonable in the parameter regime in which classical physics applies, where one can perform a binary-search type approach using a set of reference mass/lengths of value $1, 1/2, 1/4,1/8,\ldots$.  These assumptions necessarily break down near atomic scales where a polynomial relationship between desired accuracy and required energy is more appropriate.

\subsection{Divisibility of Material}
Both of the matrix multiplication constructions presented below involve some property of the system scaling inversely with the size of the instance.  For the integer matrix multiplication algorithm of Section~\ref{sec:real_mat_mult}, we assume that some material can be divided into quantities of size $1/n$.  In the Boolean matrix multiplication construction of Section~\ref{sec:bool_mat_mult}, we assume that the velocity of some components of the system can be $1/n$.  This inverse scaling breaks down at atomic scales, which is the main limit on the size of the instances for which such systems could be practically realized.  Though, as discussed at the end of Section~\ref{sec:real_mat_mult}, in an optical implementation of our integer multiplication algorithm, we would expect the roughly quadratic time and energy scaling to hold up until impressively large problem instances.\footnote{We note that, at least for multiplying square matrices, allowing properties to scale with $o(1/n)$ does not seem to help.  For other problems, such as $k$-sum, allowing material to be divisible into quantities of size $1/n^k$ can likely be leveraged. That said, such an assumption quickly becomes unrealistic---even for modest values of $k$ this assumption becomes practically unreasonable for quite modest values of $n$.}  

The specific assumption we require for our integer multiplication algorithm is that with time and energy $O( n \polylog n),$ one can construct a ``device'' with the property that if one ``pours in''  one unit of ``material'' (e.g. water, sand, light) at one end, after time $O(n)$,  $1/n \pm o(1/n^2)$ material will exit each of $n$ equally-spaced ``endpoints''.  Additionally, the amount of energy required by this system to perform such a division is either negligible, or at most $\polylog n$.  A plausible construction of such a gadget would be a binary tree of ``tubing''  through which material can flow under the force of gravity, with $n$ leaves and ``splitters'' at each of the internal nodes/junctions that divide the material flow (nearly) equally along the two downstream paths.  The construction of Section~\ref{sec:real_mat_mult} is described in terms of such a gadget.

\subsection{Classical Mechanics}
Both algorithms assume that objects operate under Newtonian mechanics: it requires a unit of energy to raise a unit mass to a height of 1 unit, and a unit mass can be moved a unit distance in a unit time, beginning and ending at rest,  requiring a unit of energy. We also assume the force of gravity acts in the usual sense.  For example, a mass at rest at the top of a length $n$ frictionless track that is at an incline of $1/n$, will take time $O(n)$ to reach the bottom.  None of our algorithms require perfectly elastic collisions, though the algorithm of Section~\ref{sec:bool_mat_mult} assumes that kinetic energy can be transferred from one object to another, losing a constant fraction (bounded below 1) of the energy.  This algorithm additionally leverages that kinetic energy scales quadratically with velocity: accelerating a unit mass object to velocity $v$ requires $O(v^2)$ energy.

\section{Integer Matrix Multiplication}\label{sec:real_mat_mult}

In this section, we consider multiplying matrices of integers. Given an $n \times n$ matrix $A$, we will construct an $O(n^2 \polylog n)$ sized physical system, taking time and energy $O(n^2 \polylog n)$, such that given a vector $b$, the matrix-vector product $Ab=c$ can be computed in time and energy $O(n \text{ } \polylog n).$  After $n$ such products, the computation corresponds to having multiplied two $n \times n$ matrices in time and energy $O(n^2 \polylog n)$.  Without loss of generality, we will assume that $A \in \{0,1\}^{n\times n}$ and $b \in \{0,1\}^n$, as the multiplication of matrices with  $r$-bit entries can trivially be reduced to $r^2$ multiplications of $\{0,1\}$ matrices.

The physical system will be constructed as a simple network of ``tubing'' and ``channels", through which a divisible ``material'' (e.g. sand, water, light) flows under the influence of gravity without friction.   We will have an array of $n$ ``channels'', with the $i$th channel corresponding to the $i$th index of the output, $c_i$.  One end of each channel will be held at one unit elevation, and the other will be held at elevation 0. The total amount of ``material'' that collects at the end of the $i$th channel will be measured, to accuracy $\ll 1/n^2$, which will be the value of $c_i$ after rounding to the nearest multiple of $1/n$.  Between each of these channels, we will also have ``garbage'' channels, whose material is never measured.

\begin{figure}[h]
    \centering
    \includegraphics[width=17cm]{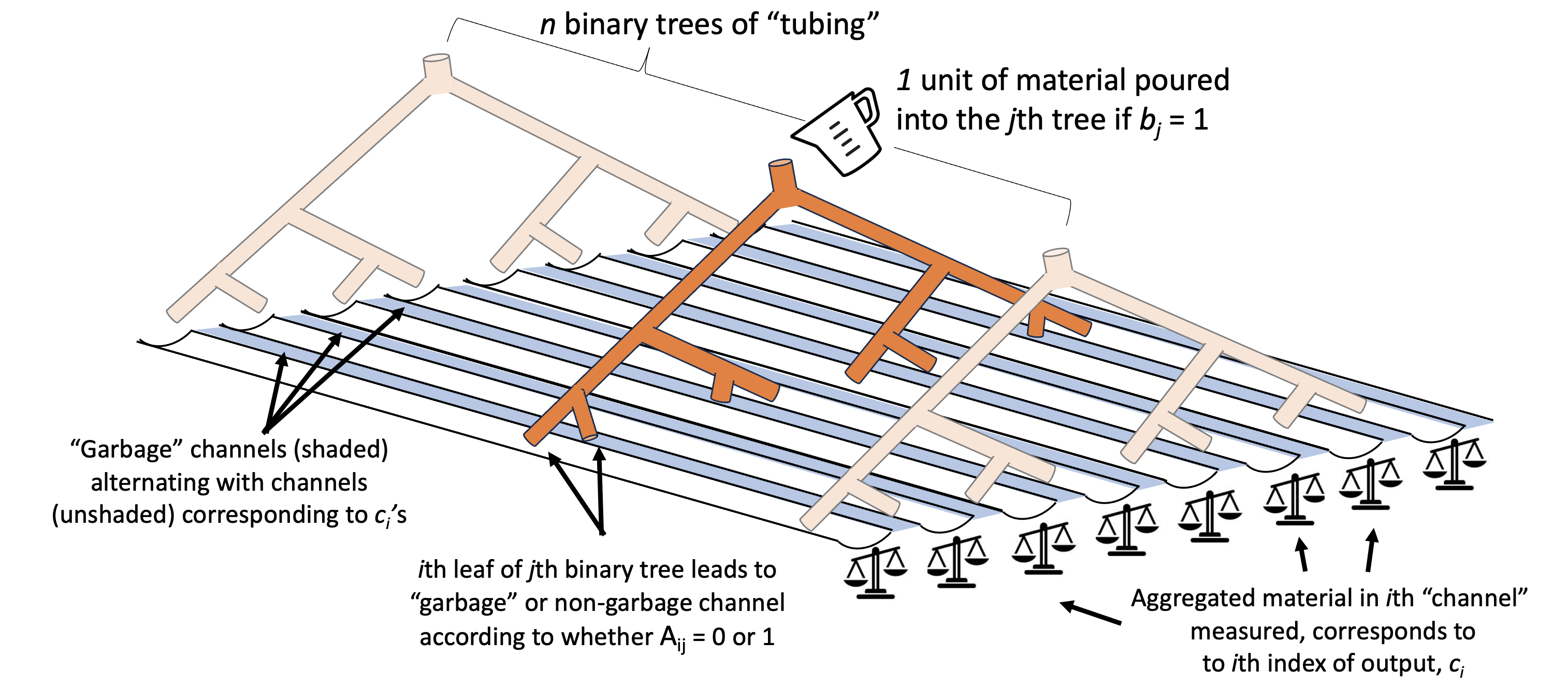}
\caption{Integer matrix multiplication: Given an $n \times n$ binary matrix, $A$, the above construction can be made in time and energy $O(n^2 \log n)$ using $O(n^2 \log n)$  material.  Once constructed, this can be used to multiply $A$ by a length $n$ binary vector, $b$ using time and energy $O(n \polylog(n)).$  Amortizing the cost of construction over $n$ such matrix vector products yields that two $n \times n$ binary matrices can be multiplied in near-quadratic time and energy.  The construction consists of $n$ binary trees of ``tubing'' through which ``material'' can flow under the influence of gravity, with the material split evenly at each node/junction.  Material from the $i$th leaf of the $j$th binary tree is directed into the $i$th channel or a ``garbage'' channel according to whether $A_{i,j}=1$ or 0.  Given the vector $b \in \{0,1\}^n,$ one unit of material is poured into the $j$th binary tree of tubing if $b_j = 1$.  The material from the $i$th channel is aggregated/measured, with the total corresponding to $(1/n)$ times the $i$th entry of the output, $c_i = \sum_{j=1}^n A_{i,j} b_j$.   Each channel has an incline of $1/n$, and the binary trees have height $O(\log n)$, sufficient for material to flow/slide in time $O(n)$ under the influence of gravity. \label{fig:realMM}}
\end{figure}

For each $j \in \{1,\ldots,n\}$, we will construct a binary tree of tubing, with $n$ ``leaves'', and height $\log n$, such that when a unit of ``material'' is input at the root, after time $O(n \log n),$ $1/n \pm 1/poly(n)$ material has come out at each ``leaf''.  This can be accomplished via ``splitters'' at each of the $O(n)$ internal nodes/junctions in the tree, each of which  splits the material equally between the two downstream paths, up to $\ll 1/n^2$ accuracy.  We assume that each of the $O(n^2)$ splitters ($O(n)$ splitters for each of the $n$ binary trees)  is an inert device that has been constructed/calibrated in time and energy $O(\log n)$.  We discuss the practical feasibility of such splitters more below.   The $j$th binary tree will be positioned $j$ units along the array of channels, such that the tubing at the $i$th leaf flows into channel $i$ if $A_{i,j}=1.$ If $A_{i,j}=0$, then the tubing at leaf $i$ of binary tree $j$ is directed towards a ``garbage'' channel.  The total size of this construction is $O(n^2 \log n),$ corresponding to $2n$ channels of length $n$ ($n$ corresponding to the outputs, and $n$ interspersed ``garbage'' channels), and $n$ binary trees each of size $O(n \log n)$.

Given this system representing matrix $A$, to multiply vector $b \in \{0,1\}^n$, for each $j \in \{1,\ldots,n\}$, we input $1$ unit (up to error $\ll 1/n^2$) of material into the $j$th binary tree of tubing if, and only if, $b_j = 1,$ and measure the amount of material that collects at each of the channels after time $O( n \log n)$; the amount of material that exits the $i$th channel, rounded to the nearest multiple of $1/n$ will be $c_i/n.$  

The correctness of the implementation is clear by construction: the amount of material entering the $i$th channel from the $j$th binary tree of tubing is $A_{i,j}b_j/n,$ and hence up to the scaling factor of $n$, the amount of material collected at the bottom of the $i$th channel is $\sum_{j=1}^n A_{i,j} b_j = c_i$.  The total energy required to perform this matrix-vector multiplication is $O(n \log n)$, corresponding to 1) lifting the $\le n$ amount of material the $O(\log n)$ distance to reach the top of the binary trees of tubing, 2) measuring each of the $\le n$ unit quantities of material to accuracy $\le 1/n^2$ to input into each of the binary trees, 3) measuring each of the $n$ outputs $c_1,\ldots,c_n$ to accuracy $\le 1/n^2$.  The total runtime is also $O(n \log n),$ consisting of 1) raising the $\le n$ units of material to height $O(\log n$, 2) the time to sequentially measure out each of the $\le n$ units of material, 3) the $O(n)$ time for the material to flow through the length $n$ tubing path and length $\le n$ channel, each of which has an incline of at least $1/n$, and 4) sequentially measuring the material emitted at each of the $n$ channels to accuracy $\ll 1/n^2$.

Finally,  this $O(n^2 \log n)$ sized construction representing matrix $A$ can be constructed in time/energy $O(n^2 \polylog n)$.  This holds assuming 1) that each of the $O(n^2)$ flow splitters---$O(n)$ per binary tree---can be calibrated to accuracy $<1/n^2$ in time/energy $O(\log n)$ per splitter, 2) that the $O(n^2 \log n)$ length of tubing can be fabricated in time/energy $O(n^2 \log n)$,  3) that the $O(n^2)$ connections between the tubing and the splitters can each be connected in time $O(\log n)$, and 4)  that the $2n$ flow channels of length $n$ can, in total, be fabricated in time $O(n^2)$.  This near-quadratic time/energy cost will be amortized across  the $n$ matrix vector multiplications, to yield an overall time/energy for multiplying two $n \times n$ matrices that is $O(n^2 \polylog n).$

\paragraph{Practical Feasibility:}  The most natural mapping of this matrix-multiplication scheme into a practically feasible construction that would have runtime and energy usage scaling nearly quadratically up to large values of $n$, would likely leverage light, rather than a material like water, or sand.  The accurate construction of the binary trees of tubing seems practically feasible given the high quality of optical beam splitters currently available.  For this application, the fact that beam splitters typically absorb (as opposed to transmit or reflect) a small constant fraction of light does not matter.  It is crucial to the construction that the beam splitters transmit and reflect nearly equal amount of light, up to error $\ll 1/n^2$---or at least that the amount of light reaching each of the $n$ leaves of each binary is equal, to this accuracy.  This property seems achievable via various $O(\log n)$ length sequences of measuring and modifying a given splitter.  Across all $O(n^2)$ splitters, the total time/energy cost would be $O(n^2 \log n)$.  As with any construction based on classical physics, this scheme is doomed to fail once $1/n$ becomes on the same scale as a single photon.  Still, this would seem to offer impressively fast and energy-efficient matrix multiplication at  large scales.

\section{Boolean Matrix Multiplication}\label{sec:bool_mat_mult}

In this section, we consider Boolean matrix multiplication---matrix multiplication of binary matrices where the elementwise product is replaced by AND, and the summation is replaced by OR: Given two $n\times n$ binary matrices, $A$, $B$, let the $n\times n$ binary matrix $C$ be defined with entry $C_{i,j} = \lor_{k=1}^n (A_{i,k} \land B_{k,j}).$  Currently, the fastest known algorithms for Boolean matrix multiplication are no better than for integer matrix multiplication.  Our main reason for describing this rather different sort of algorithm is to impress the point that it is not all that difficult to come up with physical algorithms that seem to achieve surprising runtimes and energy usages.  The algorithm of the previous section certainly seems more amenable to practical implementations than what will be described in this section.

As in Section~\ref{sec:real_mat_mult}, we will construct an $O(n^2 \polylog n)$ sized physical system that represents matrix $A$.  This construction will take $O(n^2\polylog n)$ time and energy.  Given this system, we will then be able to evaluate $Ab$ for any vector, $b$, in near linear time and energy. To motivate our algorithm, we begin with a naive approach to designing an efficient RAM algorithm for this problem:
\begin{itemize}
\item{Represent each \emph{column} of $A$ via a linked list storing the indices of the entries that are 1.  Let $L_i$ denote the list corresponding to the $i$th column.}
\item{For $j=1,\ldots,n$ we compute the (boolean) product between matrix $A$ and the $j$th column of $B$:
     \begin{itemize}
        \item initialize $C_{1,j},\ldots,C_{n,j}$ to zero.
        \item{For $k=1\ldots n$, 
        \begin{itemize}
            \item If $B_{k,j}=1$, step through $L_k$ and for each  entry $i$ (corresponding to $A_{i,k}=1$) do the following:
            \begin{enumerate}
                \item Set $C_{i,j}=1.$
                \item Remove value $i$ from all lists $L_{k'}$ for ${k'>k}$.
            \end{enumerate}            
        \end{itemize}}     
        \item Reset the lists $L_1,\ldots,L_n$ so that $L_k$ represents the nonzero indices of the $k$th column of $A$. (i.e., undo the ``removals'' of Step 2.)
    \end{itemize}}
    \end{itemize}
The above algorithm is trivially correct.  Furthermore, when processing each column of $B$, steps 1 and 2 are only ever executed once per nonzero entry of column $C_{*,j}$.  Hence each of the $n$ steps of the FOR loop would take time $O(n)$, yielding a total runtime for the matrix multiplication of $O(n^2)$, if the following held: 1) Step 2 could be accomplished in constant time (as opposed to near linear time that would be yielded by doing a binary search within each of the lists $L_{k'}$), and 2) the final step of the algorithm that resets all lists after each matrix-vector product, could be accomplished in $O(n)$ time per reset, as opposed to the $O(n^2)$ time it would take to naively rebuild all the lists.

We now describe a physical implementation of this algorithm that can be implemented in $\tilde{O}(n^2)$ time and energy.  The crux of the construction is that we will perform Step 2 using $O(\log n)$ energy in such a way that removing value $i$ from the $k'$th list will take $O(k'-k)$ time, and will only be completed as we begin to process the $k'$th entry of $B_{*,j}.$  Phrased differently, in Step 2, we need to remove $i$ from all subsequent lists $k'>k$.  However, in our implementation we will have $O(k'-k)$ time before $i$ must be removed from list $L_{k'},$ and hence we will be able to clear it very slowly over $\approx k'-k$ timesteps.  Although we have not yet described how this will be implemented, based on the kinetic energy scaling with the square of velocity it should now be plausible that the energy required would be only $O(1/(k'-k)^2)$.  Summing this energy over all $k'>k$ is a most $\sum_{i\ge 1}1/i^2 = \pi^2/6,$ as opposed to the linear energy that would be required in a RAM implementation.  We note that even if the energy required to clear a single entry in time $t$ scaled as $1/t$ instead of the optimistic $1/t^2$ scaling suggested by kinetic energy, one could still plausibly implement this high-level strategy with an energy cost of at most $\sum_{i=1}^n \frac{1}{i} = \log n$ to clear a value from all lists, affecting the total energy cost by at most logarithmic factors.

\subsection{Physical implementation} We will make an $n\times n$ physical system representing matrix $A$ on an $n\times n$ friction-less grid.  Each of the $n^2$ cells of the grid will correspond to the analogous entries of matrix $A$: we represent $A_{i,j}=0$ versus $A_{i,j}=1$ via a unit mass block being on the left size of the cell versus on the right size of the cell.   We will furthermore assume that each cell is set up in such a way that given  $c \le 1$ units of energy, it transitions from the ``1'' state to the ``0'' state in time $O(1/c)$.  This could be physically realized by imparting $\Theta(c)$ kinetic energy to the unit mass block, corresponding to velocity $\Theta(\sqrt{c})$, which would allow the block to traverse the unit length in time $O(1/\sqrt{c})\le O(1/c),$ and then coming to rest via a perfectly inelastic collision or any other way of losing its kinetic energy and reaching a configuration from where it can go back to state ``1" when necessary.

To multiply by the $j$th column of $B$, $B_{*,j}$,  for each $k$ for which $B_{k,j} = 1$, we will have a unit-mass ``agent'' which will move at unit velocity along the right side of the $k$th column corresponding to $A$.  This is analogous to traversing a linked list representing the location of the ones in the $k$th column of $A$, in the sense that the agent will only expend energy when it collides with a unit mass---namely when it arrives at an entry $A_{i,k}=1$---otherwise it continues its frictionless motion unimpeded.  

Upon colliding with a unit mass at the $i$th location while traversing the $k$th column the agent will expend $\polylog n$ energy to accomplish the following steps, corresponding to Steps 1 and 2 of the naive RAM approach: 
\begin{enumerate}
    \item Set the corresponding entry of the answer $C_{i,j}=1$.  (This could be accomplished via Newtonian mechanics by having a special frictionless track along each row of the construction, with the track of the $i$th row leading to the $i$th answer register.  An agent will send a unit mass block at unit velocity along this track, and the answer register will update from 0 to 1 upon receiving such a unit of energy.)
    \item Clear the remainder of the $i$th row, that is, for each $k'>k,$ set the entry corresponding to $A_{i,k'}$ to zero.  To accomplish this, the agent will use $O(\log n)$ energy (which can be stored at the cell itself), transferring $\approx 1/(k'-k)$ energy to the cell corresponding to $A_{i,k'}$ in time $O(k'-k)$, for all $k'>k$.  We discuss how this can be implemented below. 
    If $A_{i,k'}=1$, the corresponding cell will use the $\approx 1/(k'-k)$ energy to set the entry to 0 in time $<k'-k$; hence the entry will be in the 0 position by the time the agent corresponding to column $k'$ visits the $i$th row.  Note that this energy $1/(k'-k)$ is quadratically more than would be sufficient to zero the entry, as energy $1/(k'-k)^2$ would be sufficient to move the unit mass block a unit distance in time $k'-k$.
    \item  Finally, the agent will use constant energy to adjust its velocity (to compensate for any slowdown required to initiate the previous two steps) so that it enters row $i+1$ at velocity 1, one timestep after it entered row $i$.  [This step is important, as we must maintain the invariant that the agents in column $i'>i$  reaches row $k$ at least $i'-i$ timesteps after the $i$th agent reached that row, to allow for Step 2 to be completed.]  
\end{enumerate}

\paragraph{Transferring Energy to Clear a Row.} There are a number of ways to implement Step 2.  One approach would be to have $\log n$ frictionless tracks associated with each row.  To clear the remainder of a row, the agent will send a unit mass block at constant speed along each track.  The $\ell$th such block will travel distance $O(2^{\ell})$ and then  partition its kinetic energy (roughly) uniformly among the $2^{\ell}$ cells of the $i$th row in columns $k+2^{\ell},\ldots, k+2^{\ell+1}$.  There are a number  constructions to accomplish this  partitioning (all resembling Rube Goldberg machines at some level).  Rather than describing one, we instead sketch a more plausible practical instantiation that leverages light.  Suppose we have $\log n$ optical channels running along each row, with the $\ell$th channel having opacity $1/2^{\ell}$---namely each cell in the $\ell$th channel absorbs a $1/2^{\ell}$ fraction of the light that enters, and allows the remaining $1-1/2^{\ell}$ fraction to pass through.  Suppose the agent at the $k$th column sends one unit of energy along each of the $\log n$ optical channels associated to the given row, and consider the energy absorbed by the cell at column $k'=k+d$.  Defining $\ell = \lceil \log d \rceil$, the energy absorbed by this cell due to just this $\ell$th channel will be at least $\frac{1}{2^{\ell}}(1-\frac{1}{2^\ell})^{d-1} \ge \frac{1}{2d}(1-\frac{1}{2^\ell})^{2^{\ell}} \ge \frac{1}{8d},$ since $(1-1/c)^c$ is monotonically increasing in $c$, and $\ell \ge 1.$  

\paragraph{Resetting Before Next Matrix-Vector Product.}  One final step of the algorithm will be to reset each of the $O(n^2)$ cells that were ``cleared'' in Step 2, and also for each of the $\le n$ cells that triggered a collision, refreshing the $O(\log n)$ energy stored at that cell.  Both of these can be  accomplished in $O(n \polylog n)$ time, using $O( n \polylog n)$ energy.  For the resetting, each cleared entry can reset in time $O(n)$ (in parallel), and hence the energy required per cell could be as low as $1/n^2$ to accelerate the unit mass to velocity $1/n$.  There are various implementations, including one in which there is a weak, restorative force for each cell representing an entry $A_{i,k}=1$.  (For example, the cell could be at a slight incline allowing a gravitational restorative force favoring the $1$ position). Such a force would be sufficient to restore the cells to their original values at a timescale of $n \log n,$ but would not have a significant effect at the timescales of each matrix-vector product.

\begin{figure}[h]
    \centering
    \includegraphics[width=8 cm]{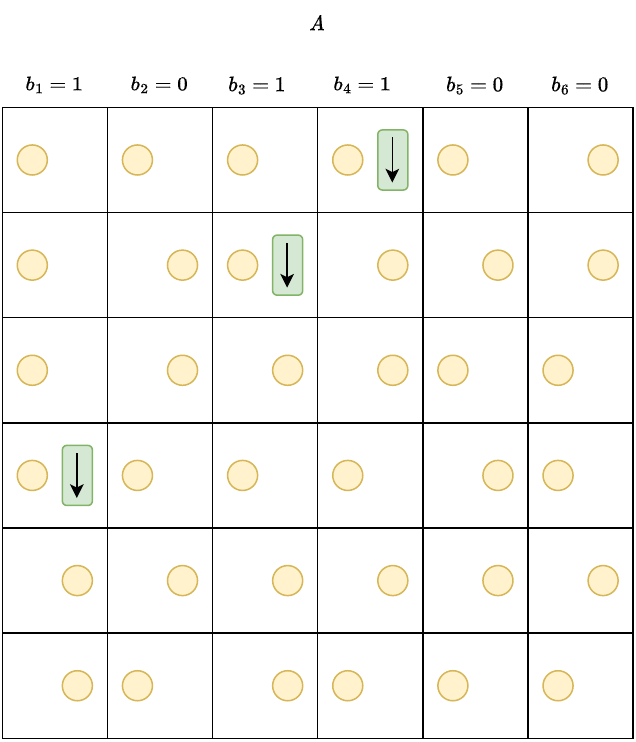}
    \caption{Boolean matrix multiplication: Each entry of matrix $A$ is encoded via an $n\times n$ grid of cells in a frictionless surface, with a unit mass in either the left or right side of cell $(i,k)$ according to $A_{i,k}=0$ or $1$.  To compute the matrix vector product $Ab$, unit mass ``agents'' (denoted as green arrows) are sent at unit velocity down each column for which $b_j = 1$.  Upon colliding with a unit mass (an entry $A_{i,k}=1$), $O(\log n)$ energy will be expended to 1) update the corresponding entry of the answer, 2) ``clear'' the rest of the $i$th row by transferring energy $\approx 1/d$ to cell $(i,k+d)$ so that the entry in cell $(i,k+d)$ can be set to 0 in time $\approx d$, before the agent in column $k+d$ arrives at cell $(i,k+d)$.  Hence each of the at most $n$ row clearing operations requires energy at most $\sum_{i=1}^n \frac{1}{i} \le \log n$.  All entries will be reset to their original positions over the course of $O(n \log n)$ timesteps using energy $O( n \log n)$ before the next matrix vector product is performed.}
    \label{fig:boolean}
\end{figure}


\section{Abstracting Physical Models of Computing}\label{sec:abstractingModels}

To simplify the design of physical algorithms, and facilitate a rigorous study of lower bounds, it would be useful to formalize an abstraction of the key computational primitives.  And ideally, this abstraction would allow the algorithm designer to work at a level removed from the minutia of exactly how and where each data element is stored and accessed.  Such an effort may be premature without a more  complete catalog of the sorts of gadgets that can be fruitfully leveraged by physical algorithms. Still, we introduce, and briefly discuss one such model. 

\paragraph{Abstracting Clock-speed/Energy Tradeoffs:}
We define the following computational model parameterized by a real number $\alpha \in [0,2]$.  The model allows for arbitrary parallelism, with processes able to create new processes, subject to the following:
\begin{itemize}
\item For a problem instance of size $n$, each process, $P$ is defined via an $O(\log n)$ size program which may include calls to create additional processes.  Process $P$ has its own \emph{rate} $r_P \ge 1$ which corresponds to the amount of time each basic operation or memory read/write takes process $P$.  Rate $r_P = 1$ corresponds to each operation taking unit time and unit energy.  A rate of $r_P = c$ corresponds to time $c$ per operation and energy usage $1/c^\alpha$. 
\item Each process requires one unit of energy to initialize.  
\item No two processes can access (read or write) the same memory location at the same time.  For example, if a process is writing a memory location at rate $r=100,$ then that memory location cannot be accessed by other processes during the 100 timesteps in which it is being written to.
\end{itemize}

Setting $\alpha = 2$ corresponds to the time/energy tradeoff in frictionless classical mechanical systems, due to kinetic energy scaling quadratically with velocity.  $\alpha = 1$ is a more modest assumption (and is presumably easier to instantiate in hardware over a larger range of problem sizes), though still  yields interesting time/energy tradeoffs.  

The following examples illustrate time/energy tradeoffs for this model.  In both cases, the algorithms are trivial---essentially naively parallelizing the task over a number of processes, all with identical rates, where the rate and level of parallelism  are jointly optimized.  The only component that requires some care is in ensuring that no two processes are reading from the same memory location at the same time.

\begin{example}[Copying a List]
Given a list of $n$ numbers to be copied, suppose we have $n^q$ processes, each running at rate $n^s$.  Each process will need to copy $n^{1-q}$ numbers, which will take time $n^{1-q+s}.$  The total energy will be the product of the number of processors and energy per processor: $n^{q}(1+n^{1-q}/{(n^s)}^{\alpha})$, where $\alpha \in [0,2]$ is the parameter governing the tradeoff between slowdown and energy usage.  For $\alpha = 1,$  this yields that for any $s\in [0,1]$ one can achieve time $O(n^{2s})$ and energy $O(n^{1-s})$---for example with $s=1/3,$ both time and energy are $O(n^{2/3}).$   For $\alpha = 2$, the analogous calculations give a tradeoff of time $O(n^{3s})$ and energy $O(n^{1-2s})$.  With $s=1/5$ both the time and energy are $O(n^{3/5})$.
\end{example}

\begin{example}[Matrix Multiplication]
Consider multiplying two $n \times n$ matrices, $A,B.$  First suppose we use $n^2$ processes, each with rate $O(n)$, where process $P_{i,j}$ is responsible for computing the $i,j$th entry of the product, $\sum_k A_{i,k}B_{k,j}$.  Since each entry of $A$ (and $B$) will be read by $n$ processes, we need to ensure that no pair of processes tries to access the same entry at the same time.  This is not difficult, and does not require any additional overhead: consider dividing time into length $n$ blocks, $[0,n],[n,2n],\ldots.$ During the $t$th block of time, let process $P_{i,j}$ read entry $A_{i,(i+j+t) \mod n}$ and $B_{(i+j+t) \mod n, j}.$  To see that no two processes are trying to access the same entry at the same time, note that the only potential collisions with process $P_{i,j}$ involve processes $P_{i',j}$ or $P_{i,j'}$.  In the case of $P_{i',j}$, a collision at time block $t$ would involve $B_{(i+j+t)\mod n, j}$ and $B_{(i'+j+t)\mod n, j}$ but these are distinct, as $i\neq i'$.   Given this lack of collisions, the runtime would be $O(n^2)$, and the energy usage would be $O\left(n^2(1+\frac{n}{n^{\alpha}})\right)=O(n^2)$ as long as $\alpha \ge 1.$   Note that in the case of $\alpha = 2,$ consistent with Newtonian mechanics, the energy overhead for initializing each process dominates the energy used in the actual computation, suggesting that subquadratic time and energy are simultaneously achievable in the $\alpha = 2$ case by using a subquadratic amount of parallelism.  Indeed, time and energy $O(n^{9/5})$ can be achieved in the $\alpha=2$ case by using $n^{9/5}$ processes, each computing $n^{1/5}$ of the entries of the product $AB$, with each process running at rate $n^{3/5}$.
\end{example}

This model suffers from some of the same drawbacks as the RAM model.  By abstracting away the details of where each bit of data is stored, for large-scale problems, the model cannot hope to realistically model the additional time/energy that must be expended by a process that needs to perform operations on bits of memory stored in ``distant'' locations.   Still, in the same way that the RAM model accurately models computations that fit on a single laptop, there is hope that future hardware could be developed that reflects the properties of the above model, at least at modest problem scales for some $\alpha>0$.

A more conceptual shortcoming of the above model is that it does not seem to be \emph{complete} in any sense.  There are properties of physical systems that can be computationally leveraged beyond the ability to reduce the energy use by slowing down a process.   In particular this model lacks the ability to aggregate values as in the algorithm for integer  matrix multiplication of Section~\ref{sec:real_mat_mult}, or the ability to average values via  diffusion.  Electromagnetic and optical phenomena are also completely absent.  Still, even within this simple and incomplete model, there might be some surprising and elegant algorithms; and there is some hope that such algorithms might be relevant to current computing settings where there is a suite of available hardware with varying speeds and energy (or monetary) costs.  Lower bounds within this restricted model might also be of interest.

\section{Concluding Thoughts}
We hope this work inspires a broader consideration of the potential landscape of time and energy requirements for problems within P, from both theoretical and practical perspectives.  Here, we  focused on matrix multiplication, leveraging Newtonian mechanics.  There are, of course, many other computational problems worth considering, and many other physical systems and forces that could be exploited for energy efficiency and parallelism, including optical phenomena,  biological processes, and gravity.  As Moore's Law wanes and alternate computing architectures are empirically investigated more fully, it may be worth developing a more complete theory of the energy or runtime gains that might be accessible via different sorts of physical systems and accompanying assumptions.

\paragraph{Acknowledgments} This research was supported by a Simons  Foundation Investigator Award.  The author would also like to thank Vijaykrishna Gurunathan for discussions on this topic.  Vijaykrishna wished to not be included as an author.  
\bibliographystyle{alpha}
\bibliography{references}

\newcommand{\etalchar}[1]{$^{#1}$}
\begin{thebibliography}{KUHK21}

\bibitem[AA11]{aaronson2011computational}
Scott Aaronson and Alex Arkhipov.
\newblock The computational complexity of linear optics.
\newblock In {\em Proceedings of the 43rd annual ACM Symposium on Theory of Computing}, pages 333--342, 2011.

\bibitem[ACNR22]{PhysRevLett.129.160502}
Simon Apers, Shantanav Chakraborty, Leonardo Novo, and J\'er\'emie Roland.
\newblock Quadratic speedup for spatial search by continuous-time quantum walk.
\newblock {\em Phys. Rev. Lett.}, 129:160502, Oct 2022.

\bibitem[Ben73]{bennett1973logical}
Charles~H Bennett.
\newblock Logical reversibility of computation.
\newblock {\em IBM Journal of Research and Development}, 17(6):525--532, 1973.

\bibitem[Ben88]{bennett1988notes}
Charles~H Bennett.
\newblock Notes on the history of reversible computation.
\newblock {\em ibm Journal of Research and Development}, 32(1):16--23, 1988.

\bibitem[Bus31]{bush1931differential}
Vannevar Bush.
\newblock The differential analyzer. a new machine for solving differential equations.
\newblock {\em Journal of the Franklin Institute}, 212(4):447--488, 1931.

\bibitem[DLMT16]{demaine2016energy}
Erik~D Demaine, Jayson Lynch, Geronimo~J Mirano, and Nirvan Tyagi.
\newblock Energy-efficient algorithms.
\newblock In {\em Proceedings of the 2016 ACM Conference on Innovations in Theoretical Computer Science}, pages 321--332, 2016.

\bibitem[GC03]{gracca2003analog}
Daniel~Silva Gra{\c{c}}a and Jos{\'e}~F{\'e}lix Costa.
\newblock Analog computers and recursive functions over the reals.
\newblock {\em Journal of Complexity}, 19(5):644--664, 2003.

\bibitem[Gro96]{grover96}
Lov~K. Grover.
\newblock A fast quantum mechanical algorithm for database search.
\newblock In {\em Proceedings of the 28th annual Symposium on the Theory of Computing}, pages 212--219, 1996.

\bibitem[HLC{\etalchar{+}}14]{hu2014memristor}
Miao Hu, Hai Li, Yiran Chen, Qing Wu, Garrett~S Rose, and Richard~W Linderman.
\newblock Memristor crossbar-based neuromorphic computing system: A case study.
\newblock {\em IEEE transactions on neural networks and learning systems}, 25(10):1864--1878, 2014.

\bibitem[HSL{\etalchar{+}}16]{hu2016dot}
Miao Hu, John~Paul Strachan, Zhiyong Li, Emmanuelle~M Grafals, Noraica Davila, Catherine Graves, Sity Lam, Ning Ge, Jianhua~Joshua Yang, and R~Stanley Williams.
\newblock Dot-product engine for neuromorphic computing: Programming 1t1m crossbar to accelerate matrix-vector multiplication.
\newblock In {\em Proceedings of the 53rd annual Design Automation Conference}, pages 1--6, 2016.

\bibitem[KUHK21]{K_ppel_2021}
Sven Köppel, Bernd Ulmann, Lars Heimann, and Dirk Killat.
\newblock Using analog computers in today{\textquotesingle}s largest computational challenges.
\newblock {\em Advances in Radio Science}, 19:105--116, dec 2021.

\bibitem[Lan61]{landauer1961irreversibility}
Rolf Landauer.
\newblock Irreversibility and heat generation in the computing process.
\newblock {\em IBM journal of research and development}, 5(3):183--191, 1961.

\bibitem[PE74]{pour1974abstract}
Marian~Boykan Pour-El.
\newblock Abstract computability and its relation to the general purpose analog computer (some connections between logic, differential equations and analog computers).
\newblock {\em Transactions of the American Mathematical Society}, 199:1--28, 1974.

\bibitem[Sha41]{shannon1941mathematical}
Claude~E Shannon.
\newblock Mathematical theory of the differential analyzer.
\newblock {\em Journal of Mathematics and Physics}, 20(1-4):337--354, 1941.

\bibitem[Sho94]{shor1994algorithms}
Peter~W Shor.
\newblock Algorithms for quantum computation: discrete logarithms and factoring.
\newblock In {\em Proceedings of the 35th annual Symposium on Foundations of Computer Science}, pages 124--134, 1994.

\bibitem[SHS{\etalchar{+}}17]{shen2017deep}
Yichen Shen, Nicholas~C Harris, Scott Skirlo, Mihika Prabhu, Tom Baehr-Jones, Michael Hochberg, Xin Sun, Shijie Zhao, Hugo Larochelle, Dirk Englund, et~al.
\newblock Deep learning with coherent nanophotonic circuits.
\newblock {\em Nature photonics}, 11(7):441--446, 2017.

\bibitem[UNT09]{uchizawa2009energy}
Kei Uchizawa, Takao Nishizeki, and Eiji Takimoto.
\newblock Energy complexity and depth of threshold circuits.
\newblock In {\em Fundamentals of Computation Theory: 17th International Symposium, FCT 2009, Wroc{\l}aw, Poland, September 2-4, 2009. Proceedings 17}, pages 335--345. Springer, 2009.

\end{thebibliography}

\end{document}